\documentclass[showpacs,preprintnumbers,amsmath,nofootinbib,aps,preprint]{revtex4}

\usepackage{epsfig,amssymb, multirow}

\begin{document}
                                                                                                                           
\title{Supersymmetry versus black holes at the LHC}

\author{Arunava Roy}
\email{arunav@olemiss.edu}
\affiliation{Department of Physics and Astronomy, University of Mississippi,
University, MS 38677-1848, USA}
                                                                                                                           
\author{Marco Cavagli\`a}
\email{cavaglia@olemiss.edu}
\affiliation{Department of Physics and Astronomy, University of Mississippi,
University, MS 38677-1848, USA}

\date{\today}

\begin{abstract}
Supersymmetry and extra dimensions are the two most promising candidates for new physics at
the TeV scale. Supersymmetric particles or extra-dimensional effects could soon be observed
at the Large Hadron Collider. We propose a simple but powerful method to discriminate the two
models: the analysis of isolated leptons with high transverse momentum. Black hole events are
simulated with the CATFISH black hole generator. Supersymmetry simulations use a combination
of PYTHIA and ISAJET, the latter providing the mass spectrum. Our results show the measure of
the dilepton invariant mass provides a strong signature to differentiate supersymmetry and
black hole events at the Large Hadron Collider. Analysis of event-shape variables and
multilepton events complement and strengthen this conclusion. 
\end{abstract}
\maketitle

\section{Introduction}
After more than a decade of intense work, CERN's Large Hadron Collider (LHC) beam
commissioning to 7 TeV is scheduled for July 2008 \cite{LHC}. The primary purpose of the
LHC is to provide experimental evidence for the Higgs particle and look for new physics
beyond the Standard Model (SM). Supersymmetry (SUSY) \cite{Martin:1997ns} is one of the best
and most studied candidates for physics beyond the SM. It provides an explanation for the
Higgs mass problem and a candidate for cold dark matter and unification of low-energy gauge
couplings by introducing superpartners to SM fields. Alternatives to SUSY include
extra-dimensional models such as large extra dimensions \cite{Arkani-Hamed:1998rs}, warped
braneworlds \cite{Randall:1999ee} or universal extra dimensions \cite{Appelquist:2000nn}. In
these models, gravity becomes strong at the TeV scale, where radiative stability is achieved.
The most astounding consequences of TeV-scale Planck mass is perhaps the production of Black
Holes (BHs) in particle colliders \cite{BHs-coll} and cosmic ray showers \cite{BHs-uhecr}. 

At a time when the LHC is becoming a reality, there is a pressing need to provide the
scientific community with tools to extract physical information from the forthcoming data. The
first step in the LHC analysis pipeline is the identification of strong experimental
signatures to distinguish the various proposals for physics beyond the SM. Comparisons of SUSY
and extra dimensions/little Higgs models have been recently investigated by various authors
\cite{Datta:2005zs,Konar:2005bd}. Here, the focus is on the difference between SUSY and BH
events. Our main result is that the measure of the dilepton invariant mass is a very strong
discriminator between SUSY and BHs. The message of this paper is thus straightforward: Take
LHC data, select dilepton events with high transverse momentum ($P_T$), measure the invariant
mass, rule out either SUSY or BHs. This procedure is discussed in detail below.

\section{Supersymmetry and Black Holes at the LHC}
According to the Minimal Supersymmetric extension of the Standard Model (MSSM)
\cite{Martin:1997ns}, all SM fermions (bosons) must have a bosonic (fermionic) partner. Since
we do not observe superpartners of SM particles at low energies, SUSY must be a broken
symmetry. A method of SUSY breaking mediated by gravitational interactions is supergravity
(SUGRA). In its minimal version, mSUGRA is determined by a point in the five-dimensional
moduli space with parameters $m_0$ (the common scalar mass at $M_{GUT}$), $m_{1/2}$ (the
common gaugino mass at $M_{GUT}$), $A_0$ (the common trilinear coupling at $M_{GUT}$),
$\tan~\beta$ (the ratio of the vacuum expectation values of the two Higgs fields), and $\mu$
(the sign of the Higgsino mass parameter). Without loss of generality, we choose $m_0=100$
GeV, $m_{1/2}=300$ GeV, $A_0=300$, $\tan\beta=2.1$ and $\mu=+1$ as SUSY benchmark when
comparing SUSY to BH events. (This is known in the literature as SUSY point A
\cite{Bartl:1996dr} or point 5 \cite{Hinchliffe:1996iu}. The choice of a different SUSY point
does not affect the results of our analysis \cite{followup}.) SUSY interactions conserve
R-parity \cite{Martin:1997ns}. R-parity conservation implies that SUSY particles are always
pair produced at the LHC and that SUSY events end with the production of a stable colorless
and chargeless lightest SUSY particle (LSP). In the analysis below, SUSY events are simulated
by generating mass spectra with ISAJET (ver.\ 7.75) \cite{Paige:2003mg} and interactions with
PYTHIA (ver.\ 6.406) \cite{Sjostrand:2006za}. All MSSM processes have been included except
Higgs production from SM interactions. 

Numerous studies have focused on BH signatures at the LHC. (See Ref.\ \cite{BH-reviews} for
reviews and details.) A quick look at BH production in colliders reveals the following. BHs
are formed when two partons interact with impact parameter less than the Schwarzschild radius
corresponding to their center-of-mass energy. The production cross-section is approximately
equal to the geometrical cross section of the event. After its formation, the BH decays
semiclassically through Hawking radiation. At the end of the Hawking decay quantum gravity
effects lead to the formation of stable remnant or the disintegration into a number of hard
quanta. Most of the particles are believed to be emitted on the brane \cite{Emparan:2000rs}.
(See also Ref.\ \cite{brane emission}.) Simulations of BH events are carried out with the
CATFISH generator \cite{Cavaglia:2006uk}. We choose a conservative benchmark model for BH
events with six extra dimensions, fundamental Planck scale 1 TeV, minimum BH mass at formation
of 2 TeV, BH mass at the end of the Hawking phase of 1 TeV and two final hard quanta
\cite{followup}.  

\section{Analysis of dilepton events}
\begin{figure}[h]
\epsfig{file=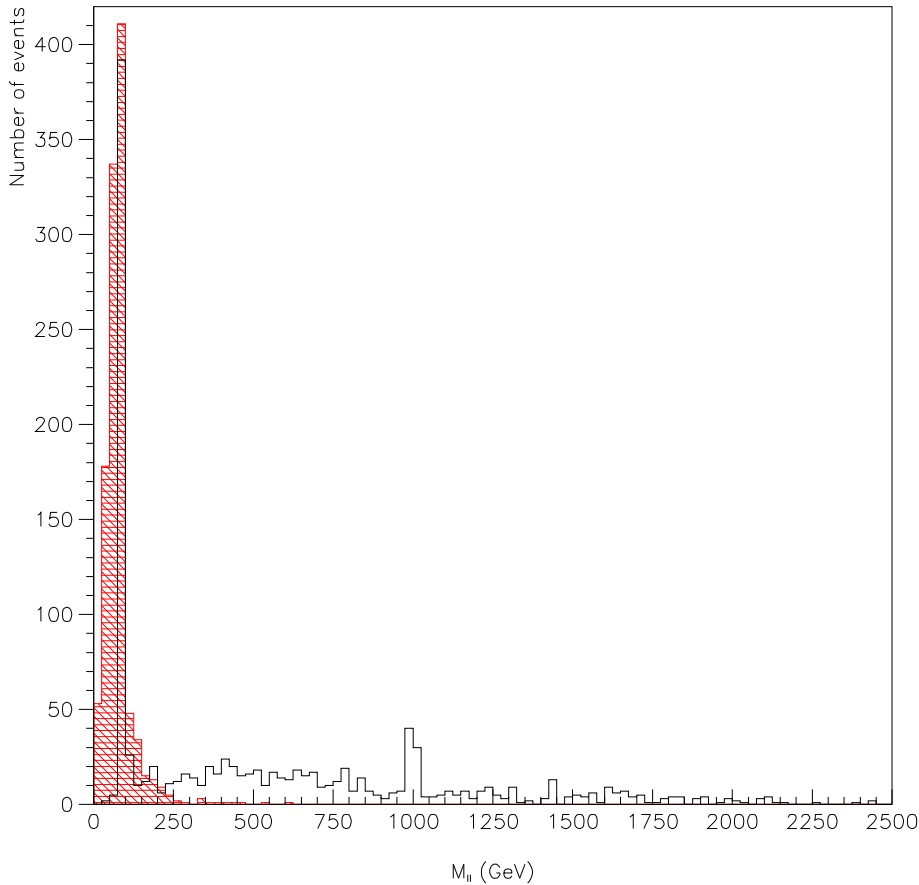,width=0.75\textwidth}
\caption{Invariant mass distribution (in GeV) for 1000 SUSY (shaded red histogram) and 1000
BH OSSF dilepton events. The SUSY distribution shows the typical endpoint due to the
presence of the LSP. The high-$P_T$ tail of the BH distribution is originated by
uncorrelated lepton pairs emitted by the BH during the Hawking evaporation phase.}
\label{fig:figure1a}
\end{figure}

The measure of the dilepton invariant mass allows the discrimination of BH and SUSY events
as follows. Our analysis is based on opposite sign, same flavor (OSSF) dileptons
\cite{Hinchliffe:1996iu}. In SUSY, the dominant source of dilepton events is the process
\begin{eqnarray*}
\tilde{\chi}_{2}^{0} \rightarrow  l^{\pm} &\tilde{l} & \\
&\downarrow &\\ 
&\tilde{l} &\rightarrow l^{\mp}~ \tilde{\chi}_{1}^{0}\,,
\label{chi2}
\end{eqnarray*}
which has a branching ratio of 27\% at LHC point 5. (The process $\tilde{\chi}_{2}^{0}
\rightarrow  \tilde{\chi}_{1}^{0} ~h$ counts for 68\%.) \cite{atlas111}. The dilepton
invariant mass is defined as 
\begin{equation}
M_{ll}=\sqrt{(E_{l^+}+E_{l^-})^2-(\textbf{p}_{l^+} +\textbf{p}_{l^-})^2}\,.
\label{Mll}
\end{equation}
Since the LSP is undetectable, the SUSY dilepton invariant mass distribution has an edge at
\cite{Hinchliffe:1996iu}
\begin{equation}
M_{ll}^{max}=
m_{\tilde{\chi}_{2}^0}~\left[\left(1-\frac{m_{\tilde{l}}^2}{m_{\tilde{\chi}_{2}^0}^2}\right)
\left(1-\frac{m_{\tilde{\chi}_{1}^0}^2}{m_{\tilde{l}}^2}\right)\right]^{\frac{1}{2}}
\sim 100\, {\rm GeV}.
\label{Mllmax}
\end{equation}
Dilepton events from BHs are not originated by a single process. Most of isolated high-$P_T$
leptons come directly from the BH, from the decay of a $Z_0$ boson, or from a top quark. (In
the latter process, the preferred channel for the dilepton event consists of one lepton from
the BH and one from a top quark.) Therefore, the dilepton invariant mass distribution
(\ref{Mll}) has no endpoint. Moreover, since the dominant decay mode of a top quark is into
hadrons \cite{PDG}, and the branching ratio of $Z_0$ into leptons is
$\Gamma(l^{+}~l^{-})/\Gamma_{tot}\sim 3.37$\% \cite{PDG}, the rate of BH OSSF dilepton events
is expected to be smaller than in SUSY. To compare the invariant mass in the two models, we
select isolated events with high $P_T$. We impose the following cuts on leptons
\cite{Hinchliffe:1996iu}:
\begin{itemize}
\item Transverse momentum $P_{T} \ge 15$ GeV;
\item Pseudorapidity \cite{Sjostrand:2006za} $\eta_l < 2.5$;
\item Isolation cut $p_T < 7$ GeV in a cone of $R=\sqrt{\Delta\eta^2+\Delta\phi^2}=0.2$,
where $\phi$ is the azimuthal angle.
\end{itemize}
Figure \ref{fig:figure1a} shows the invariant mass distribution for 1000 SUSY and BH OSSF
dilepton events. (The rate of BH-to-SUSY dilepton events at fixed luminosity is about 1:5.)
As was expected, the SUSY invariant mass distribution shows a sharp edge at $\sim$ 100 GeV.
The BH invariant mass distribution shows a peak at $\sim 90$ GeV, a second smaller peak at 1
TeV and a tail at high $P_T$. The first peak is due to dilepton events from single $Z_0$
bosons which are directly emitted by the BH. This is the dominant channel of OSSF dilepton
production in BH events. The peak at 1 TeV is due to dileptons emitted at the end of the
Hawking phase \cite{Cavaglia:2006uk}. The BH mass and the number of final hard quanta at the
end of the Hawking phase have been chosen to be 1 TeV and 2, respectively. Since the BH at the
end of the Hawking phase is expected to be electrically neutral, isolated dilepton events can
occur, for example, when the two final quanta are opposite sign leptons or a $t\bar t$ pair.
This peak is expected to be smeared out in a more realistic description of the final BH phase
\cite{followup}. The high-$M_{ll}$ tail of the distribution is originated by pairs of
uncorrelated leptons from the BH.

\begin{figure}[h]
\epsfig{file=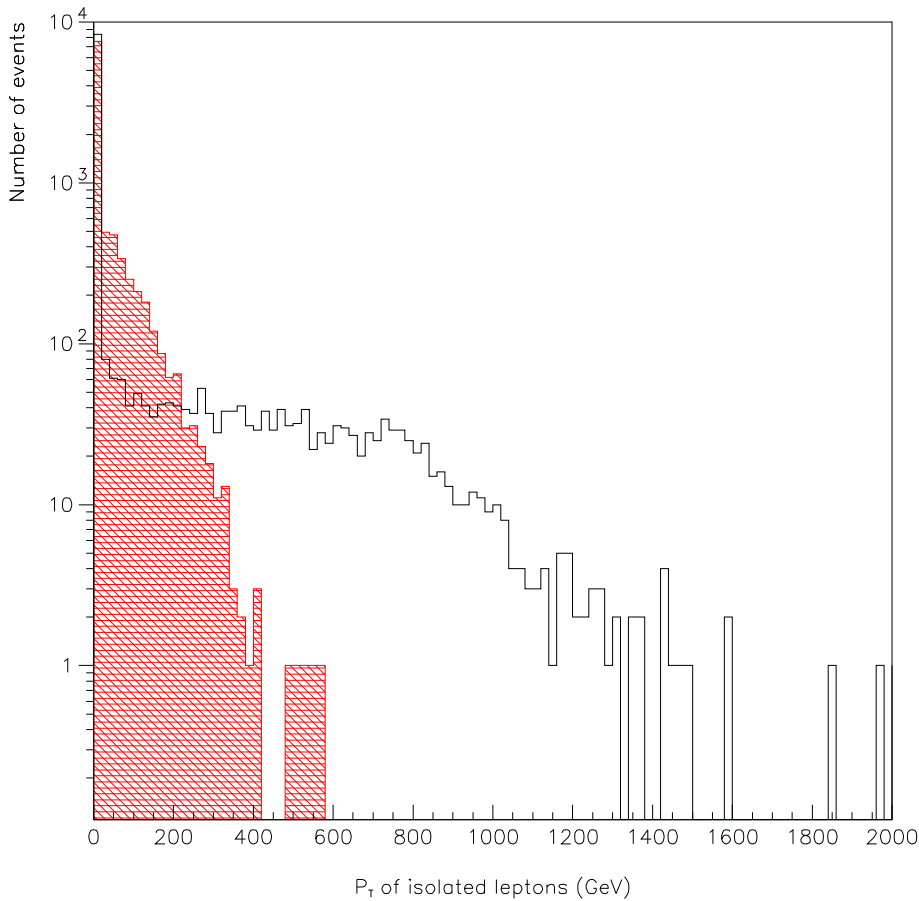,width=0.75\textwidth}
\caption{Transverse momentum distribution (in GeV) for 10,000 isolated high-$P_T$ leptons in
SUSY (shaded red histogram) and BH models. High-$P_T$ isolated leptons from BH events are
mostly emitted by the BH during the Hawking phase and have large transverse momentum.}
\label{fig:figure1b}
\end{figure}

The distribution of high-$P_T$ isolated leptons can also be used to discriminate SUSY and BH
events (Fig.~\ref{fig:figure1b}). Isolated leptons which are emitted from the BH have higher
$P_T$ than SUSY leptons. Simulations show that a BH with mass $M_{BH}\sim~{\rm few}~\times$
TeV emits few quanta during the Hawking evaporation phase, with an average energy of $E\sim
M_{BH}/{\rm few}\sim$ TeV \cite{Cavaglia:2006uk}.

\begin{figure}[h]
\epsfig{file=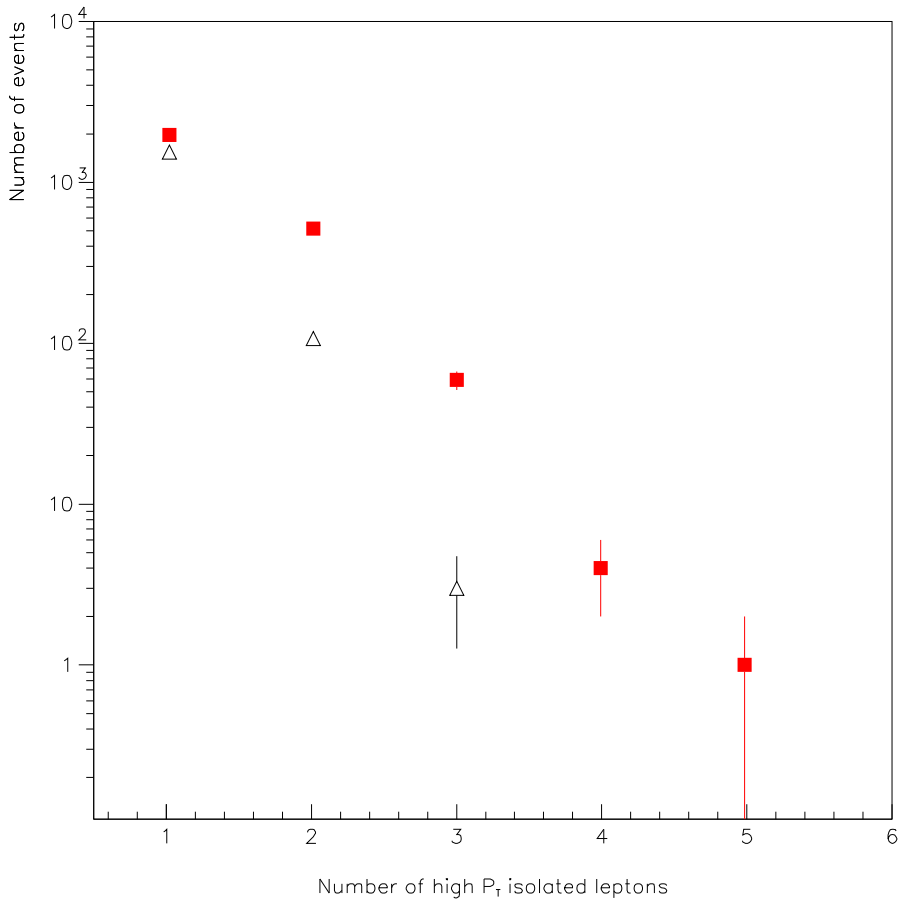,width=0.75\textwidth}
\caption{Number of high-$P_T$ leptons for SUSY and BH models (10,000 events). The number of BH
events (black open triangles) with three isolated leptons is smaller than the number of SUSY
events (red filled squares) by a factor $\sim 20$. The probability of producing BH events with
four or more leptons is virtually zero.}
\label{fig:figure2}
\end{figure}

\begin{figure}[h]
\epsfig{file=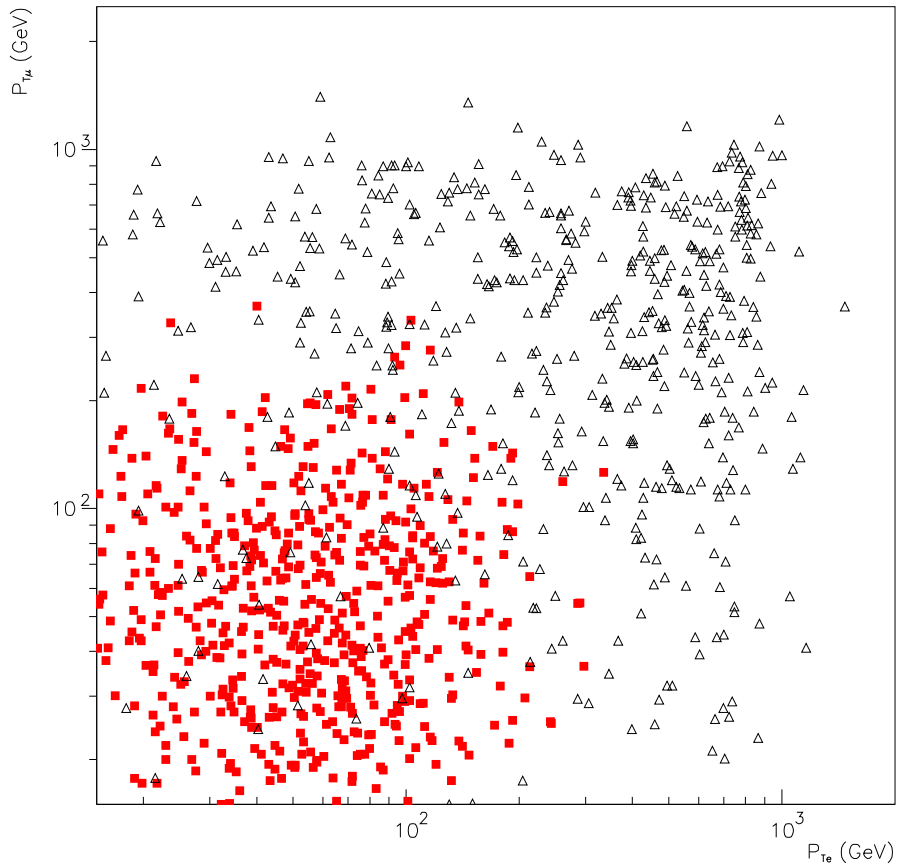,width=0.75\textwidth}
\caption{Scatter plot of transverse momentum for approximately 500 isolated opposite-flavor
dilepton events ($P_{Te}$ vs.\ $P_{T\mu}$) for SUSY (red filled squares) and BH (black open
triangles). Leptons in BH events are characterized by larger value and larger spread of their
transverse momentum.}
\label{fig:figure3}
\end{figure}

Another powerful discriminator is counting the number of isolated, high-$P_T$ leptons (of any
flavor). Figure \ref{fig:figure2} shows that SUSY events are capable of producing up to five
isolated leptons from the cascade decay of heavy sparticles. Presence of two
$\tilde{\chi}_{2}^{0}$'s in an event can produce four isolated leptons and  three leptons can
be produced by a $\tilde{\chi}_{1}^{\pm} \tilde{\chi}_{2}^{0}$ event \cite{Baer:1995tb}.
Multilepton events are rare in BH decays, where the probability of emission of more than three
isolated leptons in the Hawking phase is essentially zero. Although isolated multilepton
events are rare in both models, there is very little background. Therefore, the number of
isolated leptons and their $P_T$ are can be successfully used as SUSY/BH discriminators.
Figure \ref{fig:figure3} shows a scatter plot of high-$P_T$ electrons ($e^-$ or $e^+$) vs.\
high-$P_T$ muons ($\mu^-$ or $\mu^+$) for SUSY and BH events. SUSY isolated leptons have on
average lower $P_T$ compared to BH isolated leptons.

Dilepton events with same sign and/or opposite flavor leptons can also be used as
discriminators. The ``democratic'' nature of BH decay makes events with same/opposite flavor
leptons roughly equally probable, whereas branching ratios of SUSY events favor same-flavor
dileptons. Table \ref{table:table1} shows the branching ratios of same-/opposite-sign,
same-/opposite-flavor isolated dilepton events for SUSY and BH processes. The dominant
channel is the OSSF channel for both models. However, SUSY and BH events can be easily
discriminated by comparing the rate of same-flavor events to the rate of opposite-flavor
events. 

\begin{table}[h]
\caption{Branching ratios of high-$P_T$ isolated dileptons for SUSY and BH models. 21,000 and
100,000 events were simulated in the two cases, respectively, yielding approximately 1000
dilepton events. OS(SS) stands for opposite(same) sign and OF(SF) denotes opposite(same)
flavor.}
\begin{tabular*}{0.50\textwidth}%
{@{\extracolsep{\fill}}|c|ccccc|}
\hline
~High $P_T$ isolated dileptons~ & $SUSY$ & \% & $BH$ & \% &\\
\hline
OSSF                   & 768 & 73  & 523   & 50  &\\
\hline
SSSF                   & 65  &  6  & 103    & 10  &\\
\hline
OSOF                   & 169  & 16  & 341   & 32  &\\
\hline  
SSOF                   & 52  & 5  & 87   & 8  &\\
\hline
\end{tabular*}
\label{table:table1}
\end{table}

\begin{figure}[h]
\epsfig{file=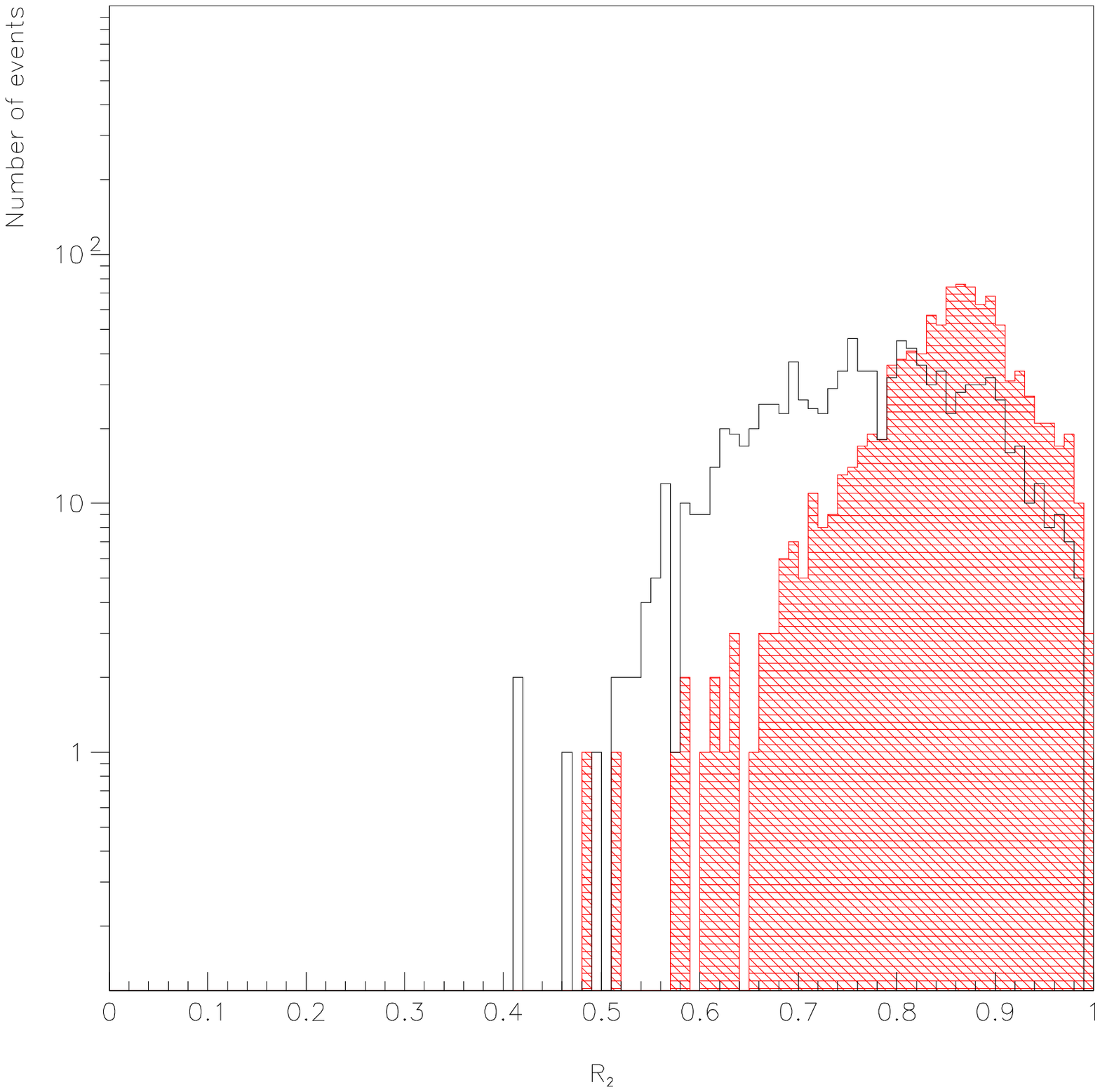,width=0.75\textwidth}
\caption{Second Fox-Wolfram moment distribution for 1000 SUSY (shaded red
histogram) and BH isolated OSSF dilepton events.}
\label{fig:figure4}
\end{figure}

Analysis of event-shape variables can be used to complement the above results. BH events are
expected to be more spherical than SUSY events because of the isotropic nature of Hawking
radiation \cite{Cavaglia:2006uk}. This statement can be made quantitative by looking at the
second Fox-Wolfram moment $R_2$ \cite{Sjostrand:2006za} (see Fig.\ \ref{fig:figure4}). SUSY
events show a sharp peak at $R_2\sim 0.85$ whereas the BH events are characterized by a
flatter distribution in the range $0.65\lesssim R_2\lesssim 0.9$. Event-shape variables
alone cannot effectively discriminate between SUSY and BHs. However, their knowledge may
prove useful when combined with the analysis of dilepton events.

\section{Conclusions}
The LHC should be able to detect SUSY or large extra dimensions, if they exist. In this
letter we have presented a powerful way to differentiate these two models based on dilepton
events. Isolated dileptons from SUSY and BH processes behave very differently. While SUSY
dileptons with high $P_T$ are characterized by a sharp endpoint of the invariant mass
distribution well below 1 TeV, BH dileptons can have invariant mass as large as several TeV.
This result can be complemented by looking at the number and flavor of isolated leptons and
event-shape variables. A simple analysis of high-$P_T$ isolated dilepton events will thus
allow the LHC to discriminate, and possibly rule out, either SUSY or BH formation at the TeV
scale. 

\noindent {\bf Acknowledgments.} We thank Vitor Cardoso, Lucien Cremaldi, Alakabha Datta and
David Sanders for fruitful discussions and are indebted to Peter Skands and Xerxes Tata for
their help with the SLHA input file and ISAJET.

\end{document}